\documentclass[preprint,floats,aps,epsfig,nofootinbib,superscriptaddress,amssymb,showpacs]{revtex4}
\usepackage{color}
\usepackage{graphicx}
\usepackage{tabularx}
\usepackage{amsmath}
\usepackage{amssymb}
\usepackage{datetime}
\usepackage[left]{lineno}
\usepackage{txfonts}
\usepackage{dcolumn}
\usepackage{bm}

\begin{document}


\title{Exploring detection of nuclearites in a large liquid
scintillator neutrino detector}

\author{Wan-Lei Guo}
\email{guowl@ihep.ac.cn} \affiliation{Institute of High Energy
Physics, Chinese Academy of Sciences, P.O. Box 918, Beijing 100049,
China}

\author{Cheng-Jun Xia}
\email{cjxia@itp.ac.cn}\affiliation{Key Laboratory of Theoretical Physics,
Institute of Theoretical Physics, Chinese Academy of Sciences, Beijing 100190, China}

\author{Tao Lin}
\email{lintao@ihep.ac.cn}\affiliation{Institute of High Energy
Physics, Chinese Academy of Sciences, P.O. Box 918, Beijing 100049,
China}

\author{Zhi-Min Wang}
\email{wangzhm@ihep.ac.cn}\affiliation{Institute of High Energy
Physics, Chinese Academy of Sciences, P.O. Box 918, Beijing 100049,
China}

\begin{abstract}

We take the JUNO experiment as an example to explore nuclearites in
the future large liquid scintillator detector. Comparing to the
previous calculations, the visible energy of nuclearites across the
liquid scintillator will be reestimated for the liquid scintillator
based detector. Then the JUNO sensitivities to the nuclearite flux
are presented. It is found that the JUNO projected sensitivities can
be better than $7.7 \times 10^{-17} {\rm cm^{-2} s^{-1} sr^{-1}}$
for the nuclearite mass $10^{15} \; {\rm GeV} \leq M \leq 10^{24}$
GeV and initial velocity $10^{-4} \leq \beta_0 \leq 10^{-1}$ with a
20 year running. Note that the JUNO will give the most stringent
limits for downgoing nuclearites with $1.6 \times 10^{13} \; {\rm
GeV} \leq M \leq 4.0 \times 10^{15}$ GeV and a typical galactic
velocity $\beta_0 = 10^{-3}$.

\end{abstract}


\maketitle

\section{Introduction}

Strange quark matter (SQM) is a hypothetical strongly interacting
matter composed of roughly equal numbers of $u$, $d$, $s$ quarks and
a small amount of electrons~\cite{Bodmer1971_PRD4-1601,
Witten1984_PRD30-272}. It is believed that SQM is the true ground
state of quantum chromodynamics, where absolutely stable SQM objects
with the baryon number $A$ ranging from that of ordinary nuclei to
neutron stars ($A\sim10^{57}$) are
expected~\cite{Xia2016_PRD93-085025}. The SQM has a little larger
density than the saturation density of ordinary nuclear matter,
which may be created in various situations, e.g., the hadronization
process of the early universe~\cite{Witten1984_PRD30-272}, collision
of binary compact stars~\cite{Madsen2002_JPG28-1737,
Bauswein2009_PRL103-011101},
type~\uppercase\expandafter{\romannumeral2} supernovae driven by
deconfinement phase transition~\cite{Vucetich1998_PRD57-5959}, and
even heavy ion collisions on Earth~\cite{Borer1994_PRL72-1415,
Weiner2006_IJMPE15-37}. The SQM objects are considered as the cold
dark matter candidates and may be presented in the cosmic radiation
reaching Earth. Light SQM objects ($A < 10^7$) are usually called
strangelets~\cite{Chin:1979yb, Farhi1984_PRD30-2379,
Berger1987_PRC35-213, Gilson1993_PRL71-332, Peng2006_PLB633-314},
while in this work we focus on heavier ones ($M > 10^{10}$ GeV)
known as nuclearites \cite{DeRujula:1984axn, DeRujula:1984hs}. Based
on their special properties, the nuclearite searches have been
performed by identifying the seismic activities with an epilinear
source on Earth \cite{Herrin2006_PRD73-043511} and the
moon~\cite{Banerdt2006_ASR37-1889}, the ionization tracks in ancient
mica \cite{Price1988_PRD38-3813} and CR39 nuclear track detectors in
the MACRO~\cite{MACRO92, MACRO00}, SLIM \cite{Cecchini:2008su}, and
Ohya \cite{Orito:1990ny} experiments, the bar excitations induced by
the thermoacoustic effect in resonant bar
detectors~\cite{Astone2013_arXiv1306.5164}, the Rutherford
backscattering of very heavy nuclei~\cite{Lv1988_PP9-385,
Liu1989_HEPNP13-103, Lowder1991_NPB24-177}, the signatures of
gravitational lensing caused by massive
nuclearites~\cite{Barnacka2012_PRD86-043001, Griest2014_ApJ786-158},
and the photons emitted when a nuclearite moves through water in the
ANTARES~\cite{Pavalas:2015nab} experiment and atmosphere in the
future JEM-EUSO~\cite{Adams:2014vgr} experiment, etc. Despite the
nonobservation of nuclearites, these experiments are able to
constrain the upper limits on the flux of cosmic nuclearites.

The liquid scintillator (LS) as the detection medium in the past
neutrino experiments has achieved great successes
\cite{Cowan:1992xc,KamLAND, Borexino, DYB}. Now the next generation
large LS detectors JUNO \cite{JUNO, CDR} and LENA  \cite{LENA} are
constructing in China and proposed in Europe, respectively. When a
nuclearite passes through the LS medium, the elastic collisions
between the nuclearite and ambient LS molecules will result in an
overheating track. Many photons from the black-body radiation of
this track can be observed by the photomultiplier tubes (PMTs).
Therefore the future large LS detectors have the ability to search
for nuclearites. A major advantage of the LS detectors is that the
LS wavelength shifters can absorb the short wavelength photons and
reemit the longer wavelength photons. This feature ensures that the
LS detectors can collect more photons from the black-body radiation
of the nuclearite track. Here we shall take the JUNO detector as an
example to explore nuclearites. The JUNO is a 20 kton multipurpose
underground LS detector and primarily determines the neutrino mass
hierarchy by detecting reactor antineutrinos.  The JUNO detector is
deployed in a 700 m underground laboratory  and consists of a
central detector, a water Cherenkov detector and a muon tracker. The
JUNO central detector holds a 20 kton LS which will be in a
spherical container with a radius of 17.7 m \cite{JUNO}. There is
1.5 m water buff region between about 18000 20-in PMTs, 36000 3-in
PMTs and the LS surface.

In this paper, we shall explore nuclearites in the JUNO LS detector
and analyze the JUNO detection capability. Comparing to the previous
calculations, the visible energy of nuclearite per unit track length
in the JUNO LS region will be reestimated in terms of the LS
fluorescence quantum yields, the PMT quantum efficiencies and the
JUNO detector design. Then we predict the JUNO sensitivities to the
nuclearite flux. In Sec. II, we outline the main features of the
nuclearite and give the maximal zenith angle below which nuclearites
may pass through the Earth rocks and reach the JUNO detector. In
Sec. III, the light yield of nuclearites traversing the JUNO LS will
be analyzed in detail. In Sec. IV, we present the JUNO sensitivity
to the nuclearite flux based on some conditions. Finally, some
discussions and conclusions will be given in Sec. V.


\section{The nuclearite energy loss}

The dominant energy loss mechanism for nuclearites passing through
matter is elastic or quasielastic collisions with the ambient atoms.
As with meteorites, the nuclearite energy loss rate can be written
as \cite{DeRujula:1984axn}
\begin{eqnarray}
\frac{d E}{d x} = -\sigma \rho \, \beta^2 \;, \label{dEdx}
\end{eqnarray}
where $\beta$ is its velocity and $\rho$ is the density of the
traversed medium. The effective nuclearite cross section $\sigma$ is
given by
\begin{eqnarray}
\sigma = \left\{
\begin{matrix}   \pi  R_0^2 & = & \pi (3 M / 4\pi \rho_N)^{2/3}\; ;   & &  M \geq
8.4 \times 10^{14} {\rm GeV}, \\   \pi {\AA}^2 & = & \pi \times
10^{-16} {\rm cm^2}\; ; & & M < 8.4 \times 10^{14} {\rm GeV},
\end{matrix} \right . \label{Sigma}
\end{eqnarray}
where the nuclearite density is estimated to be $\rho_N = 3.6 \times
10^{14} {\rm g^{-1} cm^{-3}}$ \cite{Chin:1979yb} and the nuclearite
radius $R_0$ can be easily induced from its mass $M$ and density
$\rho_N$. When the nuclearite radius $R_0 < 1 {\AA}$ ($M < 8.4
\times 10^{14} {\rm GeV}$), $\sigma$ is dominated by the nuclearite
electron atmosphere which is never smaller than the typical atomic
size with the radius of $\sim 1\ {\AA}$  \cite{DeRujula:1984axn,
Bakari:2000wa}. It is worthwhile to stress that the right-hand side
of Eq. (\ref{dEdx}) should be replaced by the constant retarding
force $- \varepsilon \sigma$ for the subsonic velocity $\beta <
\beta_c = \sqrt{\varepsilon/\rho}$ with a structural energy density
$\varepsilon \approx 10^9$ erg cm$^{-3}$ \cite{DeRujula:1984axn}.

\begin{figure}[htb]
\begin{center}
\includegraphics[scale=0.6]{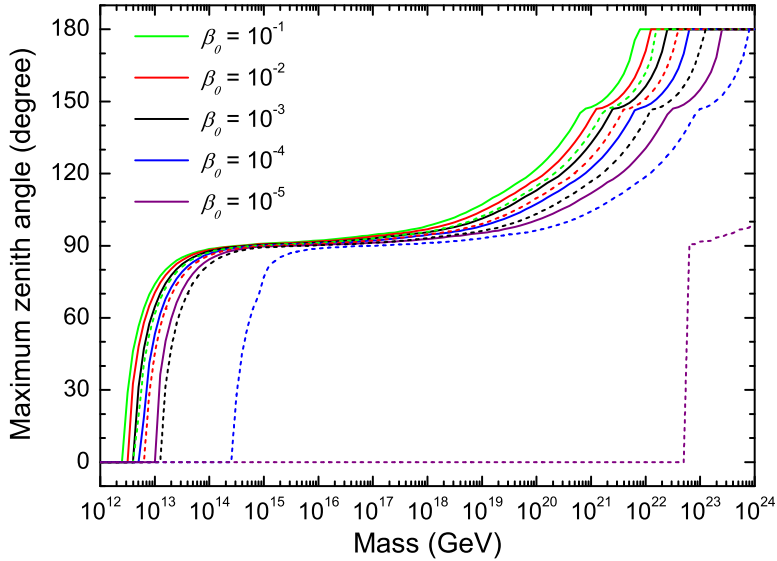}
\end{center}
\vspace{-1.0cm}\caption{ The maximal zenith angles $\theta_{\rm
max}$ below which the local velocity of nuclearite satisfies
$\beta_1 >0$ (solid lines) and $\beta_1 > \beta_{\rm min}$ (dashed
lines) for 5 typical initial velocities $\beta_0$ at the ground
level.  In Sec. \ref{S4}, we shall describe the minimal local
velocity $\beta_{\rm min}$. } \label{angle}
\end{figure}

Based on Eqs. (\ref{dEdx}) and (\ref{Sigma}), the travel length of a
nuclearite depends on its mass $M$, velocity $\beta$ and the medium
density $\rho$. Some nuclearites may pass through the Earth rocks
and arrive at the JUNO detector which has a 700 m rock overburden.
For the JUNO detectable nuclearites, they will traverse different
thicknesses of rocks based on their direction (zenith angle
$\theta_z$). In addition, one should consider the change of the
Earth matter density. In terms of the PREM Earth density profile
\cite{PREM}, we numerically calculate the maximal zenith angle
$\theta_{\rm max}$ below which nuclearites may reach the JUNO
detector, namely its local velocity $\beta_1 >0$ at the detector
level. The corresponding results have been plotted in Fig.
\ref{angle} for $10^{12} \; {\rm GeV} \leq M \leq 10^{24}$ GeV and
five typical initial velocities $\beta_0$ at the ground level. It is
clear that nuclearites from the $\theta_z = 0^\circ$ direction can
reach the JUNO detector for $M > 10^{13}$ GeV and $10^{-5} \leq
\beta_0 \leq 10^{-1}$. For a typical galactic velocity $\beta_0 =
10^{-3}$, all directional nuclearites can arrive at the detector
when $M > 2.5 \times 10^{22}$ GeV. Since the Earth density sharply
changes between the core and mantle, we can see the knee points at
$\theta_z = 146.9^\circ$ in Fig. \ref{angle}.

\section{The visible energy of nuclearite in JUNO LS \label{S3}}

When a nuclearite traverses the JUNO LS medium, the LS molecules
(${\rm C_{18}H_{30}}$) along the nuclearite path will disintegrate
into their constituents because of the nuclearite elastic or
quasielastic collisions. These heated atoms will further collide
with the ambient LS molecules and generate a hot plasma shockwave
\cite{DeRujula:1984hs}. The evolutions of effective temperature
$T(t)$ and radius $R(t)$ of the expanding thermal shockwave can be
written as \cite{DeRujula:1984axn}
\begin{eqnarray}
R^2(t) & = & \sqrt{8} \beta_1 t R_0 \;,  \label{R} \\
T(t) & = &  m \beta_1 R_0 / ( \sqrt{8} n t ) \;,  \label{T}
\end{eqnarray}
where $m$ is the LS molecule mass and $n=48$ is the number of
submolecular species. Here $\beta_1$ denotes the local velocity of
nuclearite at the detector level. The expanding cylindrical thermal
shockwave can emit lights through the black-body radiation. The
corresponding power spectrum is given by
\begin{eqnarray}
\frac{d p}{d \omega d a} = \frac{\hbar \omega^3}{4 \pi^2 c^2}
\frac{1}{e^{\hbar \omega/k T} -1} \;,
\end{eqnarray}
where $\omega$ is the angular frequency and $a$ denotes the area of
shockwave. Then we can deduce the emitted photon numbers per unit
track length $dN_\gamma/dx$ from the expanding cylindrical shock:
\begin{eqnarray}
\frac{d N_\gamma}{d x} & = & \int d\omega \int dt \; 2 \pi R(t)
\frac{d p}{d \omega d a} \frac{1}{\omega}  \nonumber \\ & = &
\frac{8^\frac{1}{4}}{2 \pi} \sqrt{\beta_1 R_0} \int d\omega \int dt
\; t^\frac{1}{2} \omega^2 \frac{1}{e^{ \omega/ T} -1} \;,
\label{dNdx}
\end{eqnarray}
where we have used the natural system of units with $\hbar = c = k
=1$. Similarly, the total emitted energy $d E_{\gamma}/d x$ can be
directly obtained through the replacement $\omega^2 \rightarrow
\omega^3$ in Eq. (\ref{dNdx}).

\begin{figure}[htb]
\begin{center}
\includegraphics[scale=0.6]{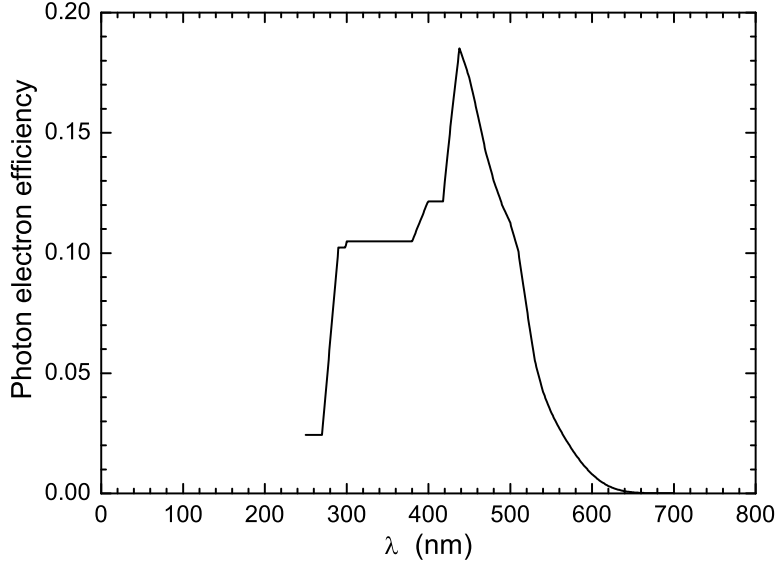}
\end{center}
\vspace{-1.0cm}\caption{ The modeled photon electron efficiency per
photon as a function of the wavelength $\lambda$. } \label{PEeff}
\end{figure}

These emitted photons from the black-body radiation cannot be
entirely detected by the JUNO PMTs since they will suffer the
absorption, reemission and Rayleigh scattering processes in the JUNO
LS \cite{CDR}. On the other hand, the PMT quantum efficiency is
related to the photon wavelength. Therefore one cannot simply use
the total emitted energy $d E_{\gamma}/d x$ or the total photon
numbers to describe the visible energy in the LS detector. It is
convenient for us to calculate the visible energy if the photon
electron (pe) efficiency per photon $\epsilon(\lambda)$ is available
for the JUNO detector. Based on the LAB, PPO and bis-MSB
fluorescence quantum yields \cite{Buck:2015jxa}, we adopt a combined
PMT quantum efficiency curve shape from the Hamamatsu PMT data
(400-800 nm) \cite{Hamamatsu} and a fixed $27\%$ efficiency (250-400
nm) to calculate $\epsilon(\lambda)$ for the wavelength range $250
{\rm \,nm} \leq \lambda \leq 800$ nm as shown in Fig. \ref{PEeff}.
We have assumed an averaged $60 \%$ survival probability of
reemitted photons from the detector center to the PMT surface and
the $75 \%$ PMT photocathode coverage. It is found that the modeled
$\epsilon(\lambda)$ approaches to zero for $ \lambda > 640$ nm. In
the absence of the related experimental data, we do not include the
contribution of the $ \lambda <250$ nm case. Note that it will not
significantly affect our final results.

\begin{figure}[htb]
\begin{center}
\includegraphics[scale=0.46]{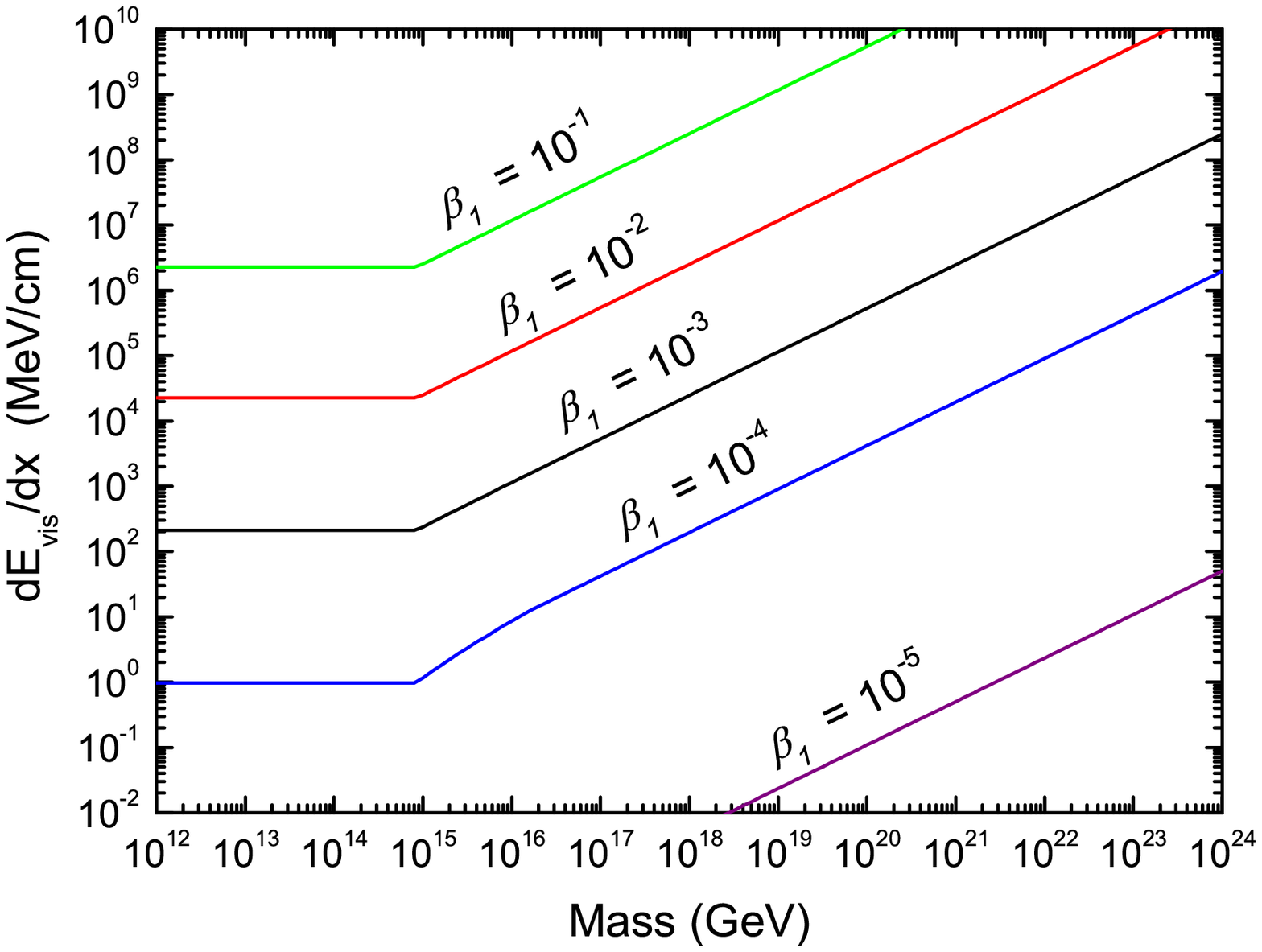}
\includegraphics[scale=0.46]{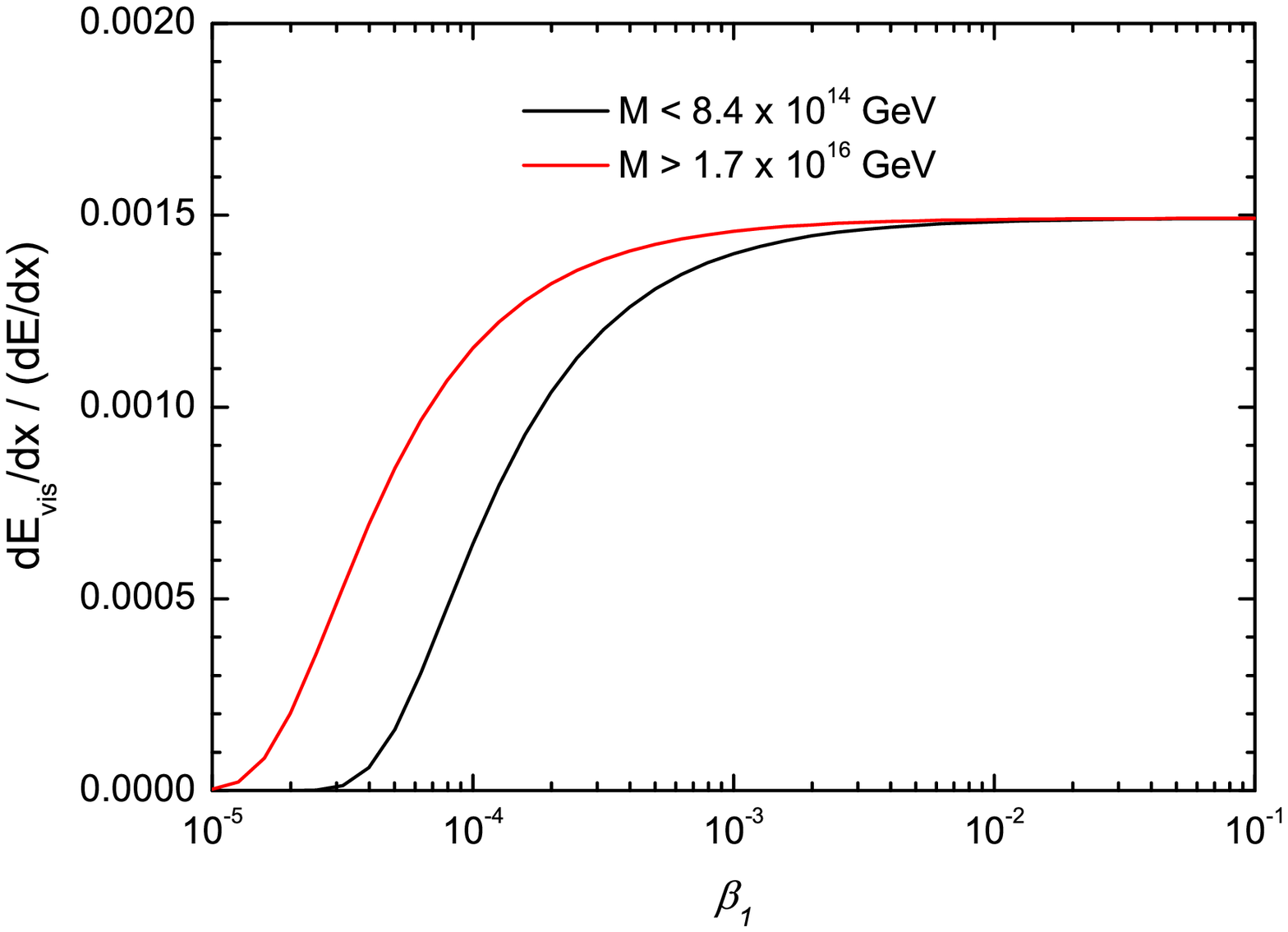}
\end{center}
\vspace{-1.0cm}\caption{ Left panel: the visible energy of
nuclearite per unit track length $d E_{vis}/d x$ as a function of
nuclearite mass for 5 typical $\beta_1$; Right panel: the ratio of
$d E_{vis}/d x$ and $d E/d x$ as a function of $\beta_1$.}
\label{energy}
\end{figure}

With the help of Eq. (\ref{dNdx}) and the photon electron efficiency
$\epsilon$ in Fig. \ref{PEeff}, we can deduce the visible energy of
nuclearite per unit track length in the JUNO LS:
\begin{eqnarray}
\frac{d E_{vis}}{d x} = \frac{ {\rm MeV}}{1200 {\rm pe}}
\frac{8^\frac{1}{4}}{2 \pi} \sqrt{\beta_1 R_0} \int d\omega \;
\omega^2 \epsilon(\omega) \int^\infty_{t_{\rm min}} dt \;
t^\frac{1}{2} \frac{1}{e^{\omega/T} -1} \;, \label{dVdx}
\end{eqnarray}
where $t_{\rm min}$ takes the larger one of $t_0 = R_0 /(\sqrt{8}
\beta_1)$ and $t_1 = (l/R_0)^2 t_0$ \cite{DeRujula:1984axn} with the
mean free path $l \approx 2.7 \AA$. In Eq. (\ref{dVdx}), we have
simply used that the 1 MeV gamma in the detector center may averagely
produce 1200 photon electrons \cite{JUNO}. Based on Eq.
(\ref{dVdx}), one may numerically calculate $d E_{vis}/d x$ as shown
in the left panel of Fig. \ref{energy}. It is clear that $d
E_{vis}/d x$ does not vary for $M < 8.4 \times 10^{14} \;{\rm GeV}$.
This is because we have adopted a constant value for
$R_0$ in Eqs. (\ref{R}) and (\ref{T}), i.e., the radius of
nuclearite electron atmosphere $1\ \AA$.
In the right panel of Fig. \ref{energy}, we plot the
ratio of $d E_{vis}/d x$ and $d E/d x$ as a function of the local
nuclearite velocity $\beta_1$. It is found that this ratio is
independent of the nuclearite mass $M$ when $M < 8.4 \times
10^{14}\; \rm GeV$ and $M > 1.7 \times 10^{16}
\; {\rm GeV}$. For the $M > 1.7 \times 10^{16}
\; {\rm GeV}$ case, namely $t_{\rm min} = t_0
> t_1$, one may easily find $d E_{vis}/d x \propto R_0^2$ through
the variable substitution $t = t' R_0$.

\section{The expected JUNO sensitivities \label{S4}}

\begin{figure}[htb]
\begin{center}
\includegraphics[scale=0.6]{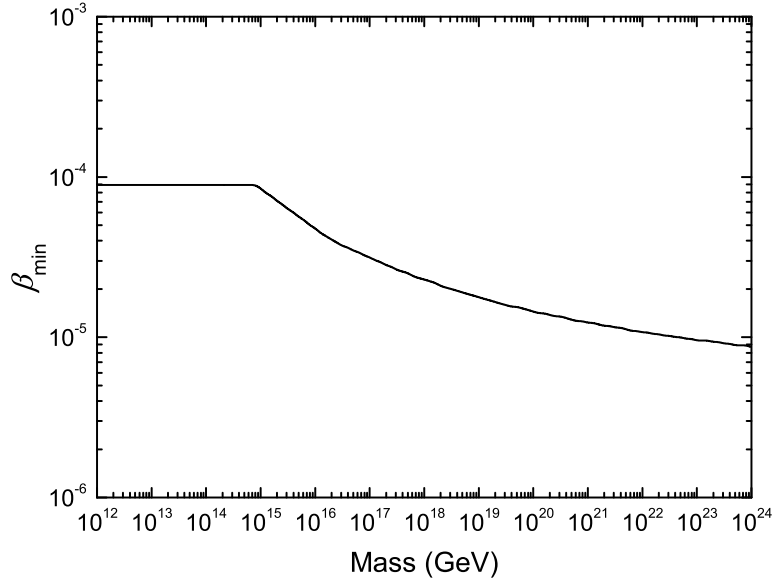}
\end{center}
\vspace{-1.0cm}\caption{The minimal local velocity $\beta_{\rm min}$
above which nuclearites can satisfy the assumed JUNO trigger. } \label{vmin}
\end{figure}

The light signals from nuclearites can be recorded when they satisfy
the JUNO trigger conditions. Here we assume that the JUNO trigger
threshold is 0.5 MeV within a $300$ ns window for the following
analyses. Then one may obtain
\begin{eqnarray}
\beta_1 \times 300 {\rm ns} \times \frac{d E_{vis}}{d x} \geq 0.5 \;
{\rm MeV} \;. \label{Cut1}
\end{eqnarray}
With the help of Eqs. (\ref{dVdx}) and (\ref{Cut1}), we calculate
the minimal local velocity $\beta_{\rm min}$ as  shown in Fig.
\ref{vmin}. It is clear that the local velocity $\beta_1$ must be
larger than $8.7 \times 10^{-6}$ for $10^{12} \; {\rm GeV} \leq M
\leq 10^{24}$ GeV. For a fixed initial velocity $\beta_0$ at the
ground level, the maximal zenith angle $\theta_{\rm max}$ can be
deduced from Eq. (\ref{dEdx}) and the requirement $\beta_1 >
\beta_{\rm min}$. In Fig. \ref{angle}, we have plotted the
corresponding $\theta_{\rm max}$ with dashed lines. It is found that
the JUNO may detect all downgoing nuclearites (zenith angle
$\theta_z < 90^\circ$) with $5.0 \times 10^{15} \; {\rm GeV} \leq M
\leq 10^{24}$ GeV and $10^{-3} \leq \beta_0 \leq 10^{-1}$. For the
$\beta_0 = 10^{-5}$ case, the downgoing nuclearites with $M
> 6.3 \times 10^{22}$ GeV can satisfy the $\beta_1
> \beta_{\rm min}$ condition.

\begin{figure}[htb]
\begin{center}
\includegraphics[scale=0.46]{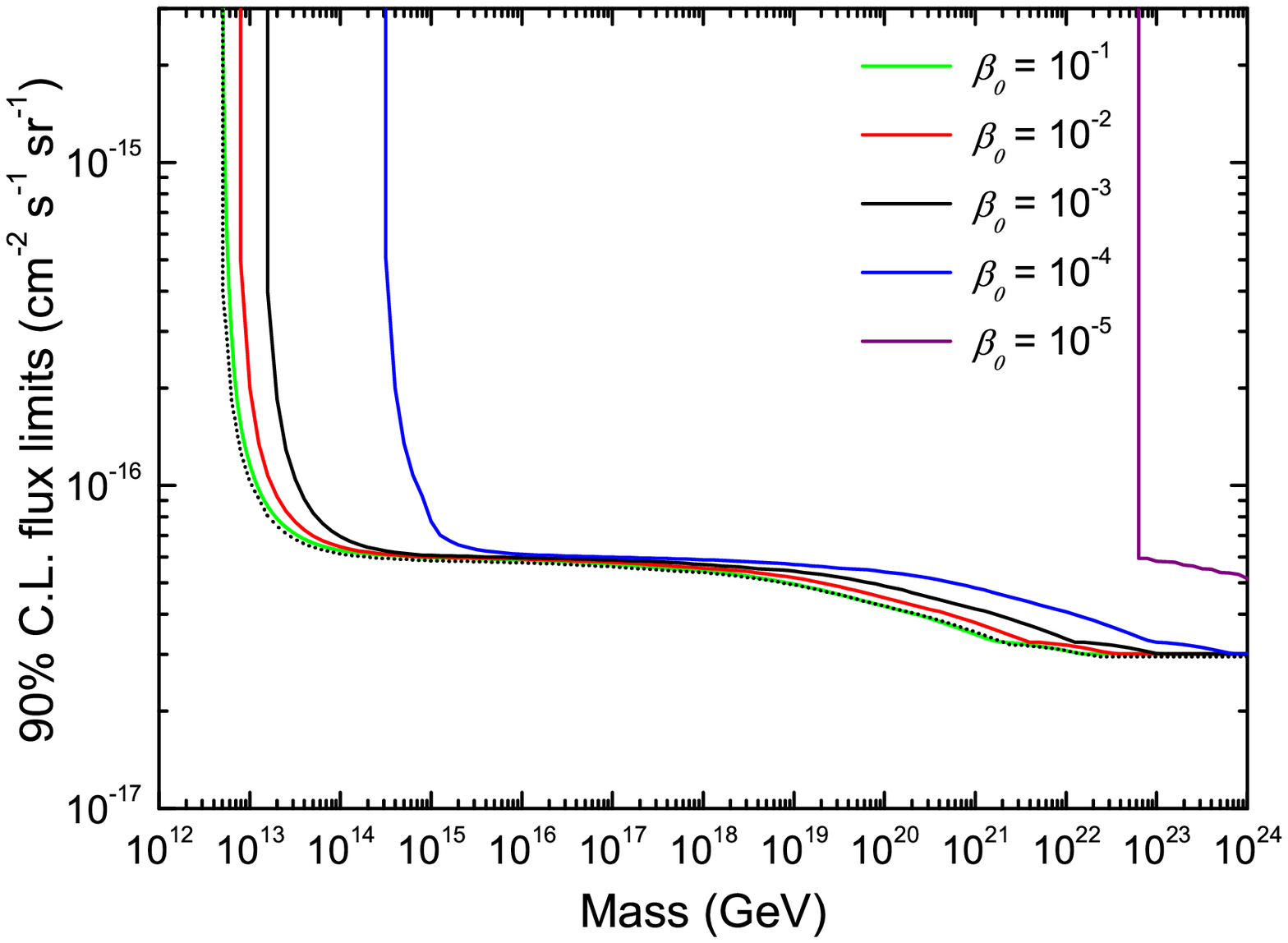}
\includegraphics[scale=0.46]{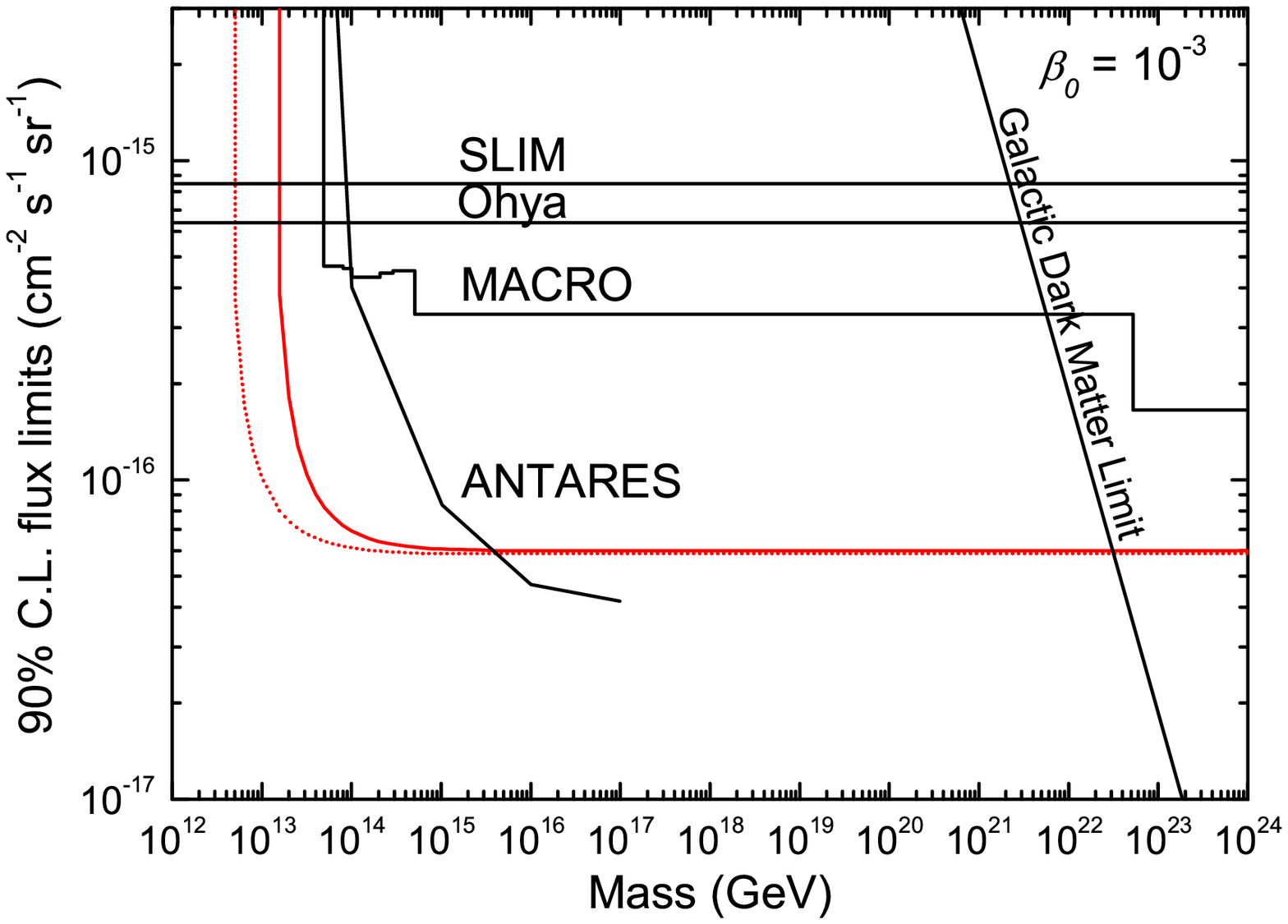}
\end{center}
\vspace{-1.0cm}\caption{The JUNO $90\%$ C.L. flux upper limits to
the all direction (left) and downgoing (right) nuclearites with the
local velocity $\beta_1 > \beta_{\rm min}$ and the effective LS
radius $R_{\rm eff} = 17.52$ m for a 20 yr running. The black and
red dotted lines describe the JUNO most optimistic limits with
$\beta_1
> 0$ and $R_{\rm eff}= 17.7$ m in the initial velocity
$\beta_0 = 10^{-3}$ case. } \label{sensitivities}
\end{figure}

The expected nuclearite numbers in JUNO can be written as
\begin{eqnarray}
N_S = 2 \pi \,  (1 -\cos \theta_{\rm max} )\; \phi \, T_{\rm run} \;
\pi \; R_{\rm eff}^2 \label{Ns}
\end{eqnarray}
where $T_{\rm run}$ is the JUNO running time and $\phi$ is the
isotropic nuclearite flux in unit of ${\rm cm^{-2} s^{-1} sr^{-1}}$.
Here we require that the nuclearite track length in LS region should
be larger than 5 m and derive the effective JUNO radius $ R_{\rm
eff}= \sqrt{{(17.7 \rm m)}^2 - (5 {\rm m}/2)^2} =17.52$ m. Then the
90\% confidence level (C.L.) upper limit $N_{90}$ to the expected
$N_{S}$ can be derived through the following formula \cite{SK, Guo}:
\begin{eqnarray}
90 \%  = \frac{\int_{N_{S} = 0}^{N_{90}} L(N_{\rm obs}|N_{S}) d N_{
S}}{\int_{N_{ S} = 0}^{\infty} L(N_{\rm obs}|N_{ S}) d N_{S}}
\label{N90}
\end{eqnarray}
with the Poisson-based likelihood function
\begin{eqnarray}
L(N_{\rm obs}|N_S) =  \frac{(N_S + N_{BG})^{N_{\rm obs}}}{N_{\rm
obs}!} e^{-(N_S + N_{BG})} \;.
\label{L}
\end{eqnarray}
To estimate the JUNO sensitivities to the nuclearite flux $\phi$, we
assume the background number $N_{BG} =0$ and take the observed event
number $N_{\rm obs} = N_{BG} = 0$ for a 20 yr running. With the help
of Eqs. (\ref{Ns}) and (\ref{N90}), we plot the  $90\%$ C.L. flux
upper limits (solid lines) for five typical initial velocities
$\beta_0$ as shown in the left panel of Fig. \ref{sensitivities}. It
is clear that the JUNO sensitivities are better than $7.7 \times
10^{-17} {\rm cm^{-2} s^{-1} sr^{-1}}$ for $10^{15} \; {\rm GeV}
\leq M \leq 10^{24}$ GeV and $10^{-4} \leq \beta_0 \leq 10^{-1}$.
The JUNO is only sensitive to a narrow parameter space of the
$\beta_0 = 10^{-5}$ case because of $\beta_1 < \beta_{\rm min}$. In
addition, the most optimistic limit (black dotted line) has been
also plotted for the $\beta_0 = 10^{-3}$ case where we only require
$\beta_1 > 0$ and take $R_{\rm eff} = 17.7$ m.

To compare with the MACRO \cite{MACRO00}, ANTARES
\cite{Pavalas:2015nab}, SLIM \cite{Cecchini:2008su} and Ohya
\cite{Orito:1990ny} experimental results, we calculate the JUNO
upper limit on the downgoing nuclearites in the $\beta_0 = 10^{-3}$
case. Our numerical results are presented in the right panel of Fig.
\ref{sensitivities}. It is clear that the JUNO sensitivity is far
better than the MACRO, SLIM and Ohya limits. Note that the JUNO will
give the most stringent limit in the range of $1.6 \times 10^{13} \;
{\rm GeV} \leq M \leq 4.0 \times 10^{15}$ GeV. In the most
optimistic case (red dotted line), the above range can be extended
to $5.0 \times 10^{12} \; {\rm GeV} \leq M \leq 4.0 \times 10^{15}$
GeV. Here we have also plotted the galactic dark matter (DM) upper
limit $\phi_{\rm max} = \rho_{\rm DM} \beta_0 / (2 \pi M)$
\cite{DeRujula:1984axn, MACRO00} where nuclearites are assumed to
contribute all of the local DM density $\rho_{\rm DM} = 0.39$ GeV
cm$^{-3}$ \cite{PDG}. For $M > 3.1 \times 10^{22}$ GeV, the galactic
DM limit is dominant.  In the future, the JEM-EUSO experiment will
give a more stringent limit $\phi < 10^{-20} {\rm cm^{-2} s^{-1}
sr^{-1}}$ for $M > 5 \times 10^{22}$ GeV \cite{Adams:2014vgr}.

\section{Discussions and Conclusions}

As mentioned in Sec. \ref{S3}, the photon electron efficiency per
photon $\epsilon(\lambda)$ is not considered for the $ \lambda <250$
nm range because of the absence of the related experimental data. If
these data are available in future, we shall derive the larger $d
E_{vis}/d x$ than those in the left panel of Fig. \ref{energy}. Then
the smaller $\beta_{\rm min}$ will be expected. The predicted
sensitivities (solid lines from $\beta_1 > \beta_{\rm min}$) in Fig.
\ref{sensitivities} will approach the most optimistic limits (the
corresponding dotted lines from $\beta_1 > 0$). It is clear that our
results do not change significantly for $M > 1.0 \times 10^{14}$ GeV
and $\beta_0 = 10^{-3}$. Note that the nuclearite mass cannot be
correctly reconstructed from the incomplete $\epsilon(\lambda)$ when
a nuclearite is really detected by the JUNO LS detector. In
addition, the JUNO can only give the mass lower bound for very large
$d E_{vis}/d x$ because of the PMT saturation.

In conclusion, we have investigated nuclearites  in the JUNO LS
detector. Comparing to the previous calculations, the visible energy
of nuclearite in the LS has been  estimated in detail. Then we give
the JUNO detectable range of the zenith angle for the nuclearite
mass $10^{12} \; {\rm GeV} \leq M \leq 10^{24}$ GeV and five typical
initial velocities $\beta_0$ at the ground level. Finally, we
present the JUNO sensitivities to the nuclearite flux for a 20 yr
running. It is found that the JUNO sensitivities to all directional
nuclearites are better than $7.7 \times 10^{-17} {\rm cm^{-2} s^{-1}
sr^{-1}}$ for $10^{15} \; {\rm GeV} \leq M \leq 10^{24}$ GeV and
$10^{-4} \leq \beta_0 \leq 10^{-1}$. For the downgoing nuclearites,
the expected sensitivities are much better than those from the
MACRO, SLIM and Ohya experiments in the case of $\beta_0 = 10^{-3}$.
Note that the JUNO will give the most stringent limits for $1.6
\times 10^{13} \; {\rm GeV} \leq M \leq 4.0 \times 10^{15}$ GeV.

\acknowledgments

We are grateful to Jun Cao, Liang Zhan and Shan-Gui Zhou for their
very useful discussions and critical remarks. This work is supported
in part by the National Nature Science Foundation of China (Grants
No.~11525524, No. 11575201 and No. 11621131001), and the Strategic
Priority Research Program of the Chinese Academy of Sciences under
Grant No. XDA10010100.


\end{document}